\documentclass[rmp,twocolumn,twoside]{revtex4}

\usepackage{graphicx}
\usepackage{multirow}
\usepackage{float}

\newcommand{\papertitle}{The NOESIS Network-Oriented Exploration, Simulation, and Induction System}

\begin{document}

\title{\papertitle}

\newcommand{\urgaffiliation}{Department of Computer Science and Artificial Intelligence, University of Granada, Spain}

\author{V\'ictor Mart\'inez}
\email{victormg@acm.org}
\affiliation{\urgaffiliation}

\author{Fernando Berzal}
\email{berzal@acm.org}
\affiliation{\urgaffiliation}

\author{Juan-Carlos Cubero}
\email{jc.cubero@decsai.ugr.es}
\affiliation{\urgaffiliation}

\begin{abstract}
Network data mining has become an important area of study due to the large number of problems it can be applied to. This paper presents NOESIS, an open source framework for network data mining that provides a large collection of network analysis techniques, including the analysis of network structural properties, community detection methods, link scoring, and link prediction, as well as network visualization algorithms. It also features a complete stand--alone graphical user interface that facilitates the use of all these techniques. The NOESIS framework has been designed using solid object--oriented design principles and structured parallel programming. As a lightweight library with minimal external dependencies and a permissive software license, NOESIS can be incorporated into other software projects. Released under a BSD license, it is available from \url{http://noesis.ikor.org}.
\end{abstract}
\maketitle

\tableofcontents

\onecolumngrid

\vspace*{1cm}

\twocolumngrid

\section{Introduction}
Data mining techniques are intended to extract information from large volumes of data \citep{tan2006introduction}. Data mining includes tasks such as classification, regression, clustering, or anomaly detection, among others. Traditional data mining techniques are typically applied to tabulated data. Novel techniques have also been devised for semi-structured or structured data, since exploiting the relationships among instances from a dataset leads to new research and development opportunities \citep{getoor2005link}. 

For example, network data mining has been used to predict previously unknown protein interactions in protein-protein interaction networks \citep{martinez2014prophnet}. It has also been used to study and predict future author collaborations and tendencies in co-authorship networks \citep{pavlov2007finding}. Different network mining techniques are used by popular internet search engines to rank the most relevant websites \citep{page1999pagerank}. These are only some examples of the large number of applications of network data mining.

There are many software tools that facilitate the analysis of networked data. Some tools provide closed solutions for end users who need to work with their own network data sets, whereas other tools cater to software developers as software libraries that collect network analysis algorithms. Most tools allow the analysis of network topology and the computation of different structural properties of networks having thousands or even millions of nodes. Some of them also include implementations of specific network data mining techniques, such as community detection algorithms or predictive models, including link prediction \citep{lu2011link} and epidemic models \citep{keeling2005networks}.

For instance, Pajek \citep{batagelj1998pajek}, NodeXL \citep{smith2009analyzing}, Gephi \citep{bastian2009gephi}, and UCINET \citep{borgatti2002ucinet} are widely used for social network analysis (SNA). Graphviz \citep{ellson2002graphviz} and Cytoscape \citep{shannon2003cytoscape} are two well--known alternatives for network visualization. Finally, igraph \citep{csardi2006igraph} and NetworkX \citep{schult2008exploring} are two popular software libraries of network algorithms. A more comprehensive and up--to--date list of available software tools can be found at Wikipedia: \url{https://en.wikipedia.org/wiki/Social_network_analysis_software}.

NOESIS, whose name stands for Network--Oriented Exploration, Simulation, and Induction System, is a software framework for network analysis and mining. It tries to combine the best features of closed social network analysis tools and extensible algorithm libraries, while providing support for parallel execution, something that most listed tools lack. NOESIS is completely written in Java and its source code has been released under a permissive BSD open source license.

Our paper is structured as follows. In Section 2, the NOESIS architectural design principles are briefly described. Section 3 covers the network analysis techniques included in NOESIS. Network data mining techniques are surveyed in Section 4. Finally, Section 5 describes the NOESIS project current status and roadmap.

\section{The design of the NOESIS framework}
NOESIS has been designed to be an easily--extensible framework whose architecture provides the basis for the implementation of network data mining techniques. In order to achieve this, NOESIS is designed around abstract interfaces and a set of core classes that provide essential functions, which allows the implementation of different features in independent components with strong cohesion and loose coupling. NOESIS components are designed to be maintainable and reusable.

\subsection{System architecture}
\begin{figure}[t]
  \centering
   \includegraphics[width=0.5\textwidth]{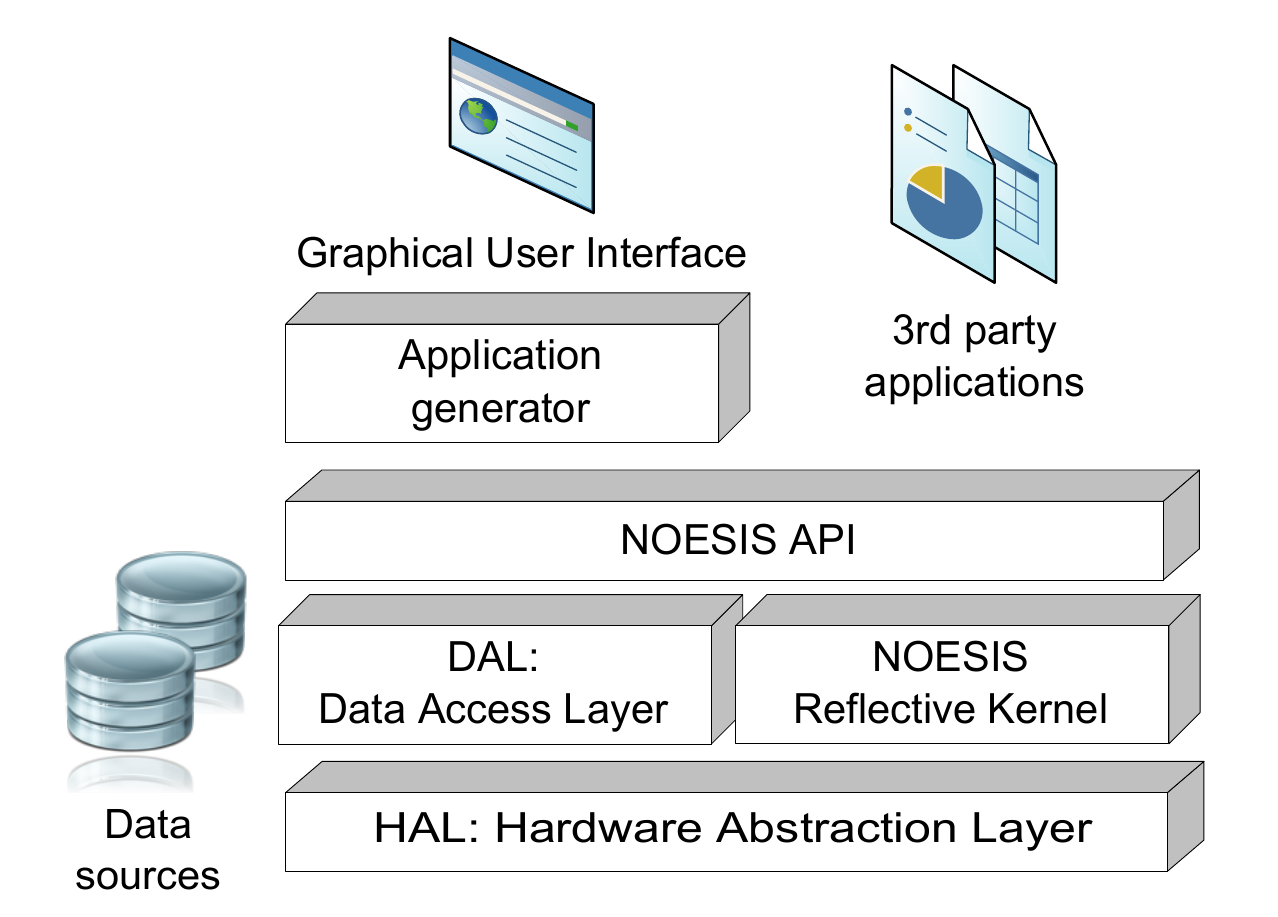}
  \caption{The NOESIS framework architecture and its core subsystems.}
  \label{fig:noesis}
 \end{figure}

The NOESIS framework architecture and its core subsystems are displayed in Figure \ref{fig:noesis}. These subsystems are described below. 

The lowest-level component is the hardware abstraction layer (HAL), which provides support for the execution of algorithms in a parallel environment and hides implementation details and much of the underlying technical complexity. This component provides different building blocks for implementing well-studied parallel programming design patterns, such as MapReduce \citep{dean2008mapreduce}. For example, we would just write \textit{result = (double) Parallel.reduce( index -\textgreater x[index] * y[index], ADD, 0, SIZE-1)} to compute the dot product of two vectors in parallel. The HAL does not only implement structured parallel programming design patterns, but it is also responsible for task scheduling and parallel execution. It allows the adjustment of parallel execution parameters, including the task scheduling algorithm.

The reflective kernel is at the core of NOESIS and provides its main features. The reflective kernel provides the base models (data structures) and tasks (algorithms) needed to perform network data mining, as well as the corresponding meta-objects and meta-models, which can be manipulated at run time. It is the underlying layer that supports a large collection of network analysis algorithms and data mining techniques, which are described in the following section. Different types of networks are dealt with using an unified interface, allowing us to choose the particular implementation that is the most adequate for the spatial and computational requirements of each application. Algorithms provided by this subsystem are built on top of the HAL building blocks, allowing the parallelized execution of algorithms whenever possible.

The data access layer (DAL) provides an unified interface to access external data sources. This subsystem allows reading and writing networks in different file formats, providing implementations for some of the most important standardized network file formats. This module also enables the development of data access components for other kinds of data sources, such as network streaming.

Finally, an application generator is used to build a complete graphical user interface following a model driven software development (MDSD) approach. This component provides a friendly user interface that allows users without programming skills to use most of the NOESIS framework features.

\subsection{Core classes}
\begin{figure}[t]
  \centering
   \includegraphics[width=0.5\textwidth]{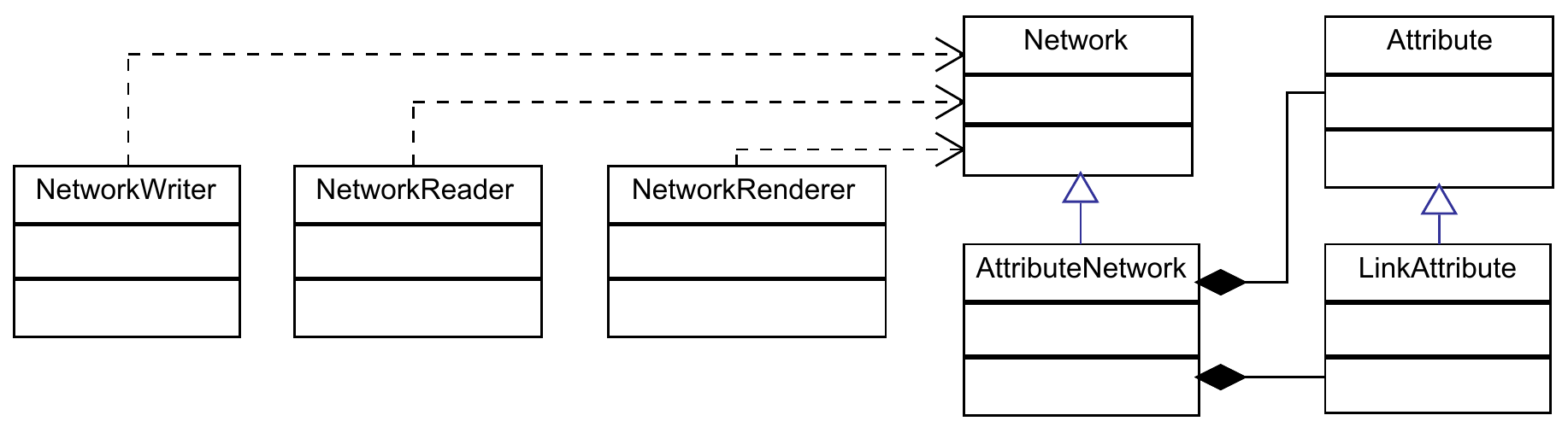}
  \caption{Some of the NOESIS core classes and interfaces represented as an UML class diagram.}
  \label{fig:uml}
 \end{figure}

The core classes and interfaces shown in Figure \ref{fig:uml} provide the foundation for the implementation of different types of networks with specific spatial and computational requirements. Basic network operations include adding and removing nodes, adding and removing links, or querying a node neighborhood. More complex operations are provided through specialized components. 

\begin{figure*}[t]
  \centering
   \includegraphics[width=\textwidth]{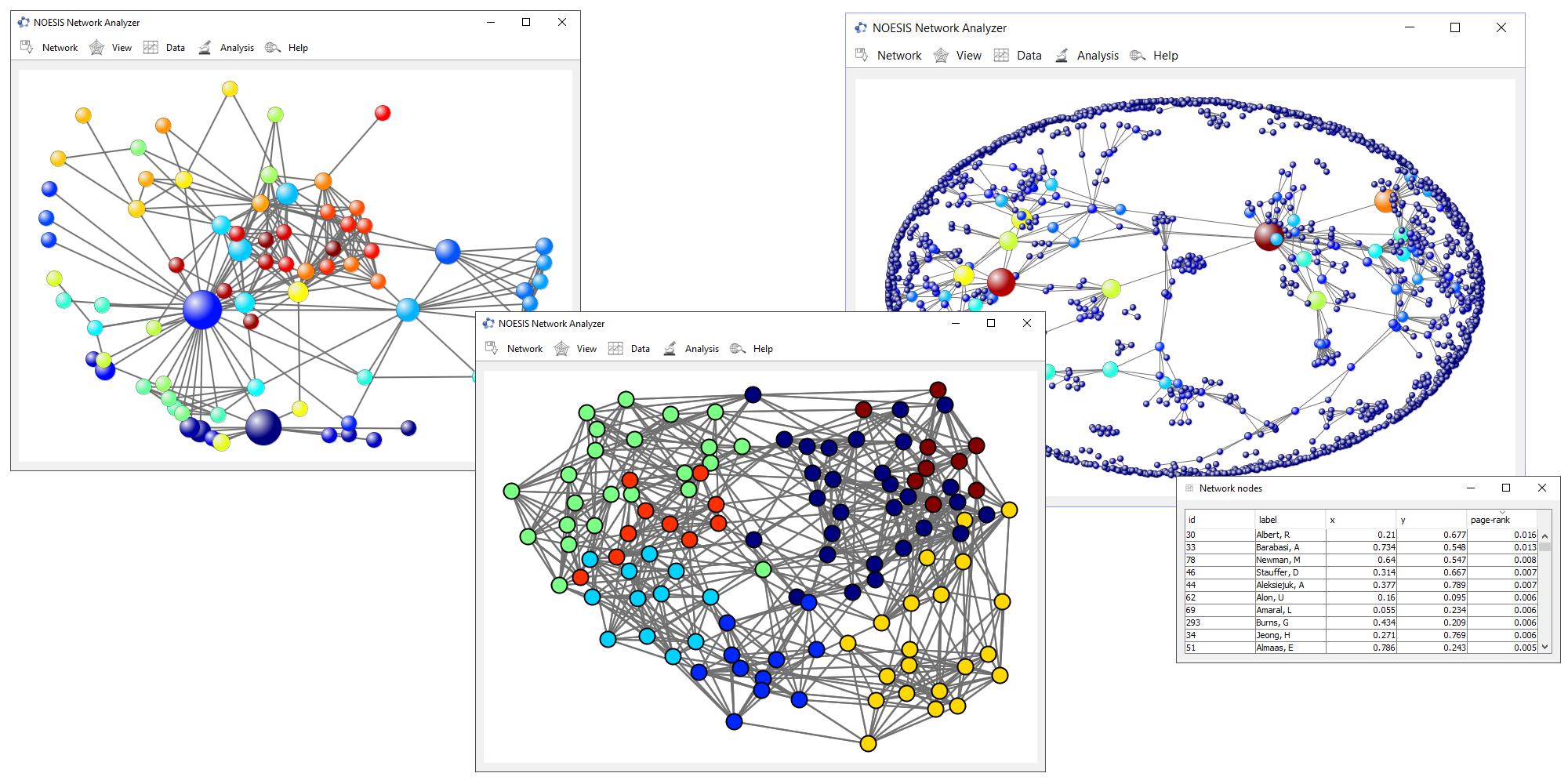}
  \caption{Different screenshots of the NOESIS graphical user interface.}
  \label{fig:noesisui}
 \end{figure*}
 
NOESIS supports networks with attributes both in their nodes and their links.  These attributes are defined according to predefined data models, including categorical and numerical values, among others.

\subsection{Supported data formats}
Different file formats have been proposed for network datasets. Some data formats are more space efficient, whereas others are more easily parseable. 

NOESIS supports reading and writing network data sets using the most common data formats. For example, the GDF file format is a CSV-like format used by some software tools such as GUESS and Gephi. It supports attributes in both nodes and links. Another supported file format is GML, which stands for Graph Modeling Language. GML is a hierarchical ASCII-based file format. GraphML is another hierarchical file format based on XML, the ubiquitous eXtensible Markup Language developed by the W3C.

Other file formats are supported by NOESIS, such as the Pajek file format, which is similar to GDF, or the file format of the datasets from the Stanford Network Analysis Platform (SNAP) \citep{snapnets2014}.

\subsection{Graphical user interface}

In order to allow users without programming knowledge to use most of the NOESIS features, a lightweight easy--to--use graphical user interface is included with the standard NOESIS framework distribution. The NOESIS GUI allows non--technical end users loading, visualizing, and analyzing their own network data sets by applying all the techniques provided with NOESIS. 

Some screenshots of this GUI are shown in Figure \ref{fig:noesisui}. A canvas is used to display the network in every moment. The network can be manipulated by clicking or dragging nodes. At the top of the window, a menu gives access to different options and data mining algorithms. The \textit{Network} menu allows loading a network from an external source and exporting the results using different file formats, as well as creating images of the current network visualization both as raster and vector graphics image. The \textit{View} menu allows the customization of the network appearance by setting specific layout algorithms and custom visualization styles. In addition, this menu allows binding the visual properties of nodes and links to their attributes. The \textit{Data} menu allows the exploration of attributes for each node and link. Finally, the \textit{Analysis} menu gives access to most of the techniques that will be described in the following sections.

\section{Network analysis tools}
NOESIS is designed to ease the implementation of network analysis tools. It also includes reusable implementations of a large collection of popular network--related techniques, from graph visualization \citep{tamassia2013handbook} and common graph algorithms, to  network structural properties \citep{newman2010networks} and network formation models \citep{jackson2008social}. The network analysis tools included in NOESIS and the modules that implement them are introduced in this section.

\begin{figure*}[t]
  \centering
   \includegraphics[width=0.3\textwidth]{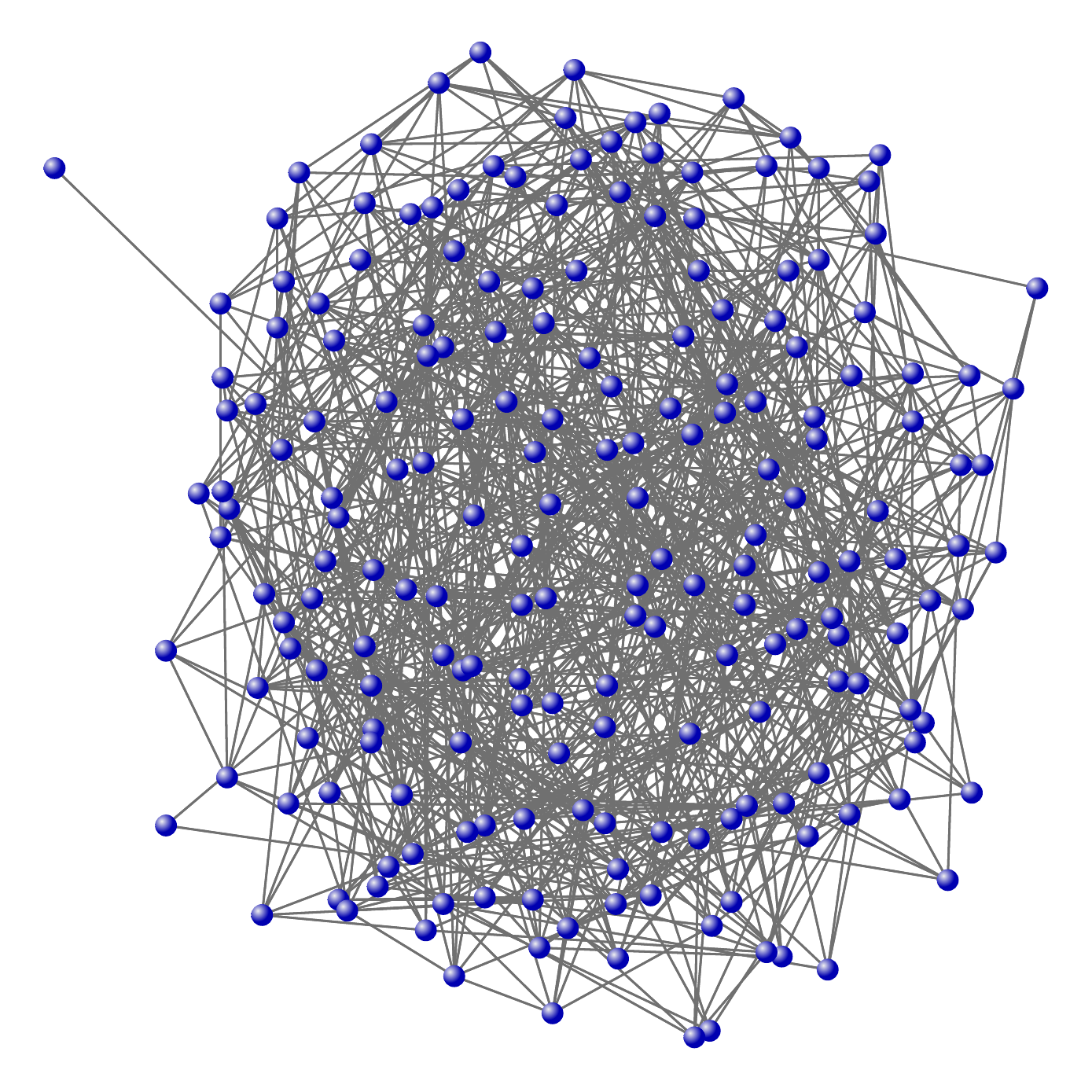}   
   \includegraphics[width=0.3\textwidth]{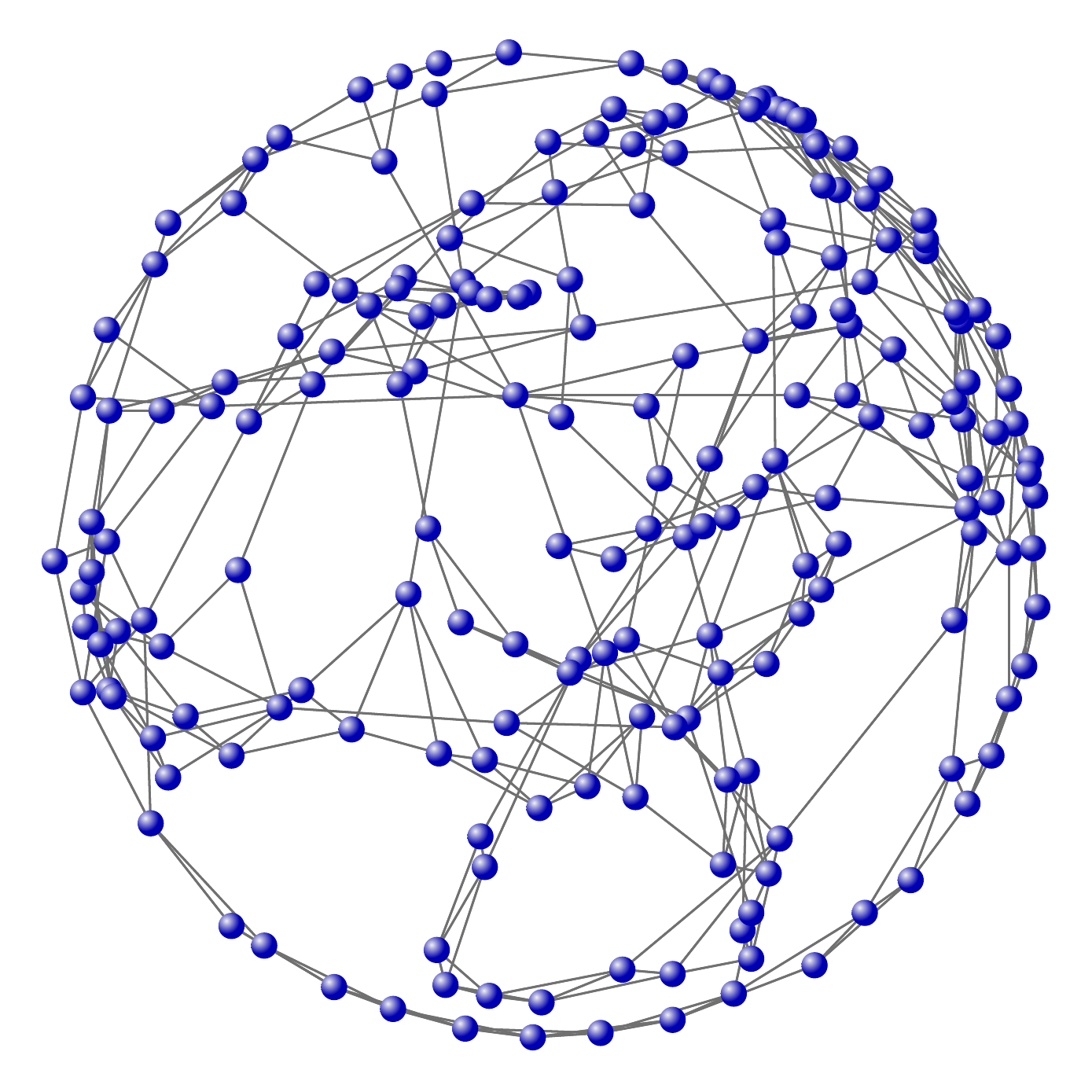}  
   \includegraphics[width=0.3\textwidth]{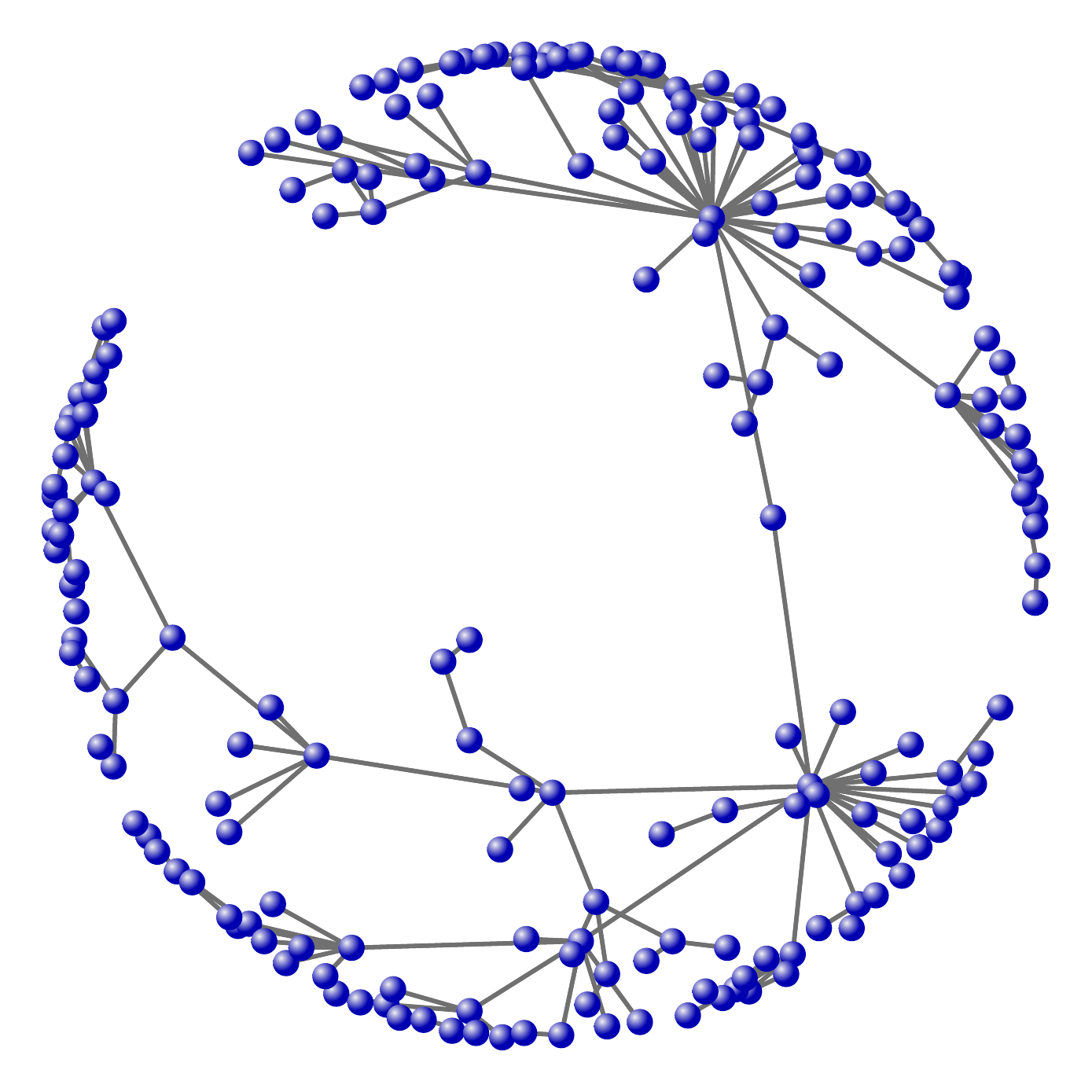} 
  \caption{Random networks generated using the Erd\"os-–R\'enyi model (left), the Watts--Strogatz model (center), and the Barab\'asi--Albert model (right).}
  \label{fig:models}
 \end{figure*}

\subsection{Network models}
NOESIS implements a number of popular random network generation models, which are described by probability distributions or random processes. Such models have been found to be useful in the study and understanding of certain properties or behaviors observed in real-world networks.

Among the models included in NOESIS, the Erd\"os-–R\'enyi model \citep{erdHos1959random} is one of the simplest ones. The Gilbert model \citep{gilbert1959random} is similar but a probability of existence is given for links instead. The anchored network model is also similar to the two previous models, with the advantage of reducing the occurrence of isolated nodes, but at the cost of being less than perfectly random. Finally, the connected random model is a variation of the anchored model that avoids isolated nodes.

Other models included in NOESIS exhibit specific properties often found in real-world networks. For example, the Watts--Strogatz model \citep{watts1998collective} generates networks with small-world properties, that is, low diameter and high clustering. This model starts by creating a ring lattice with a given number of nodes and a given mean degree, where each node is connected to its nearest neighbors on both sides. In the following steps, each link is rewired to a new target node with a given probability, avoiding self-loops and link duplication.

Despite the small-world properties exhibited by networks generated by the Watts--Strogatz model are closer to real world networks than those generated by models based on the Erd\"os-–R\'enyi approach, they still lack some important properties observed in real networks. The Barab\'asi--Albert model \citep{albert2002statistical} is another well-known model that generates networks whose node degree distribution follows a power law, which leads to scale-free networks. This model is driven by a preferential attachment process, where new nodes are added and connected to existing nodes with a probability proportional to their current degree. Another model with very similar properties to Barab\'asi--Albert's model is the Price's citation model \citep{newman2003structure}.

In addition to random network models, a number of regular network models are included in NOESIS. These models generate synthetic networks that are useful in the process of testing new algorithms. The networks regular models include complete networks, where all nodes are interconnected; star networks, where all nodes are connected to a single hub node; ring networks, where each node is connected to its closest two neighbors along a ring; tandem networks, like ring model but without closing the loop; mesh network, where nodes are arranged in rows and columns, and connected only to their adjacent nodes; toruses, meshes where nodes in the extremes of the mesh are connected; hypercubes; binary trees; and isolates, a network without links.

\subsection{Structural properties of networks}
Network structural properties allow the quantification of features or behaviors present in the network. They can be used, for instance, to measure network robustness or reveal important nodes and links. NOESIS considers three types of structural properties: node properties, node pair properties (for pairs both with and without links among them), and global properties.

NOESIS provides a large number of techniques for analyzing network structural properties. Many structural properties can be computed for nodes. For example, in-degree and out-degree, indicate the number of incoming and outgoing links, respectively. Related to node degree, two techniques to measure node degree assortativity have been included: biased \citep{piraveenan2008local} and unbiased \citep{piraveenan2010classifying} node degree assortativity. Node assortativity is a score between $-1$ and $1$ that measures the degree correlation between pairs of connected nodes. The clustering coefficient can also be computed for nodes. The clustering coefficient of a node is the fraction of its neighbors that are also connected among them.

Reachability scores are centrality measures that allow the analysis of how easy it is to reach a node from other nodes. The eccentricity of a node is defined as the maximum distance to any other node \citep{hage1995eccentricity}. The closeness, however, is the inverse of the sum of the distance from a given node to all others \citep{bavelas1950communication}. An adjusted closeness value that normalizes the closeness according to the number of reachable nodes can also be used. Inversely to closeness, average path length is defined as the mean distance of all shortest paths to any other node. Decay is yet another reachability score, computed as the summation of a delta factor powered by the path length to any other node \citep{jackson2008social}. It is interesting to note that with a delta factor close to $0$, the measure becomes the degree of the node, whereas with a delta close to $1$, the measure becomes the component size of the component the node is located at. A normalized decay score is also available.

Betweenness, as reachability, is another way to measure node centrality. Betweenness, also known as Freeman's betweenness, is a score computed as the count of shortest paths the node is involved in \citep{freeman1977set}. Since this score ranges from $2n-1$ to $n^2-(n-1)$ for $n$ the number of nodes in strongly-connected networks, a normalized variant is typically used.

Finally influence algorithms provide a different perspective on node centrality. These techniques measure the power of each node to affect others. The most popular influence algorithm is PageRank \citep{page1999pagerank}, since it is used by the Google search engine. PageRank computes a probability distribution based on the likelihood of reaching a node starting from any other node. The algorithm works by iteratively updating node probability based on direct neighbors probabilities, which leads to convergence if the network satisfies certain properties. A similar algorithm is HITS \citep{kleinberg1999authoritative}, which stands for hyperlink-induced topic search. It follows an iterative approach, as PageRank, but computes two scores per node: the hub, which is a score related to how many nodes a particular node links, and the authority, which is a score related to how many hubs link a particular node. Both scores are connected by an iterative updating process: authority is updated according to the hub scores of nodes connected by incoming links and hub is updated according to authority scores of nodes connected by outgoing links. Eigenvector centrality is another iterative method closely related to PageRank, where nodes are assigned a centrality score based on the summation of the centrality of their neighbors nodes. Katz centrality considers all possible paths, but penalizes long ones using a given damping factor \citep{katz1953new}. Finally, diffusion centrality \citep{kang2012diffusion} is another influence algorithm based on Katz centrality. The main difference is that, while Katz considers infinite length paths, diffusion centrality considers only paths of a given limited length.

In the following example, we show how to load a network from a data file and compute its structural properties using NOESIS, its PageRank scores in particular:
\begin{verbatim}
FileReader fileReader =
    new FileReader("karate.gml");
NetworkReader reader =
    new GMLNetworkReader(fileReader);
Network network = reader.read();
PageRank task = new PageRank(network);
NodeScore score = task.call();
\end{verbatim}

Different structural properties for links can also be computed by NOESIS. For example, link betweenness, which is the count of shortest paths the link is involved in, or link rays, which is the number of possible paths between two nodes that cross a given link. Some of these properties are used by different network data mining algorithms.

\subsection{Network visualization techniques}
\begin{figure*}[t]
  \centering
   \includegraphics[width=0.45\textwidth]{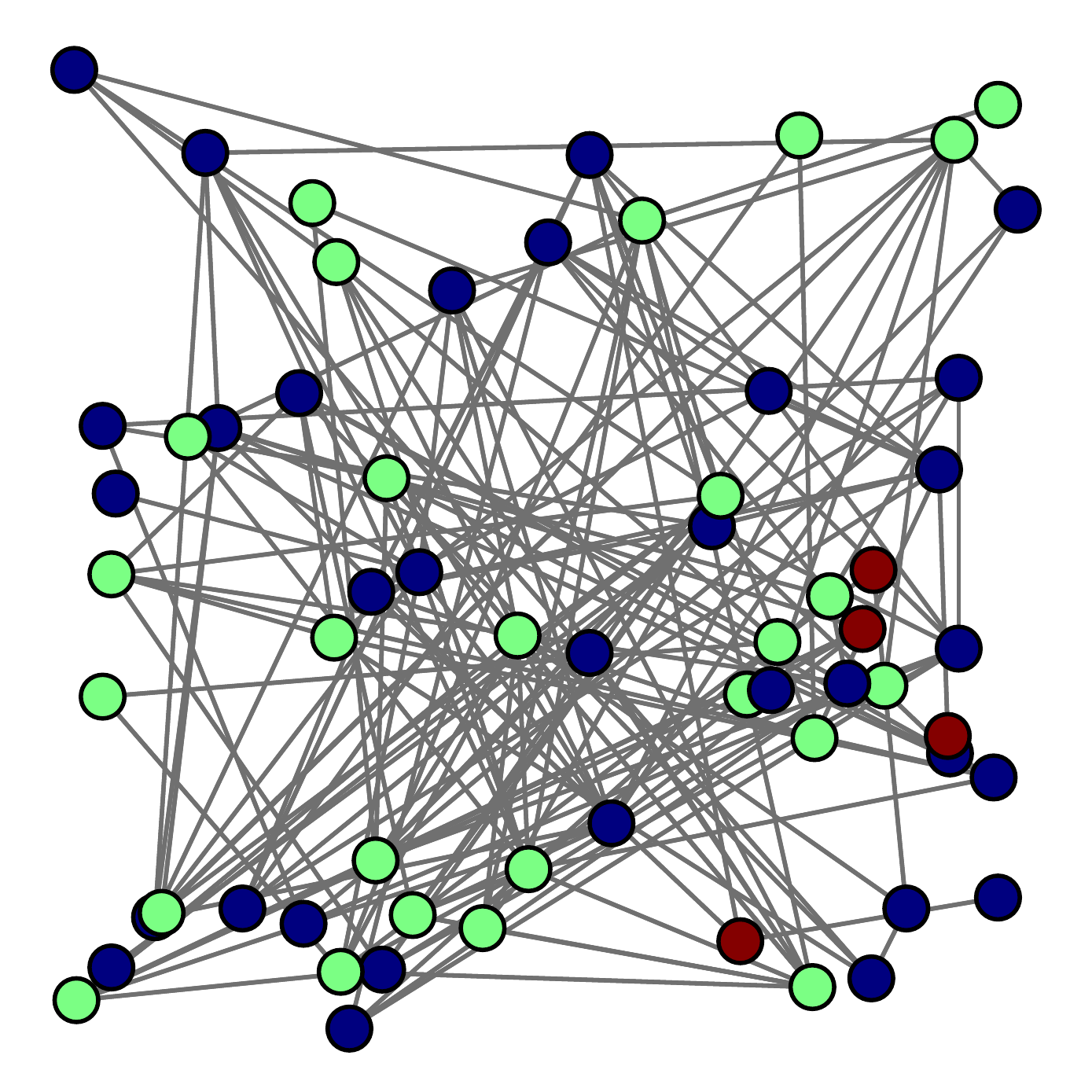}
   \includegraphics[width=0.45\textwidth]{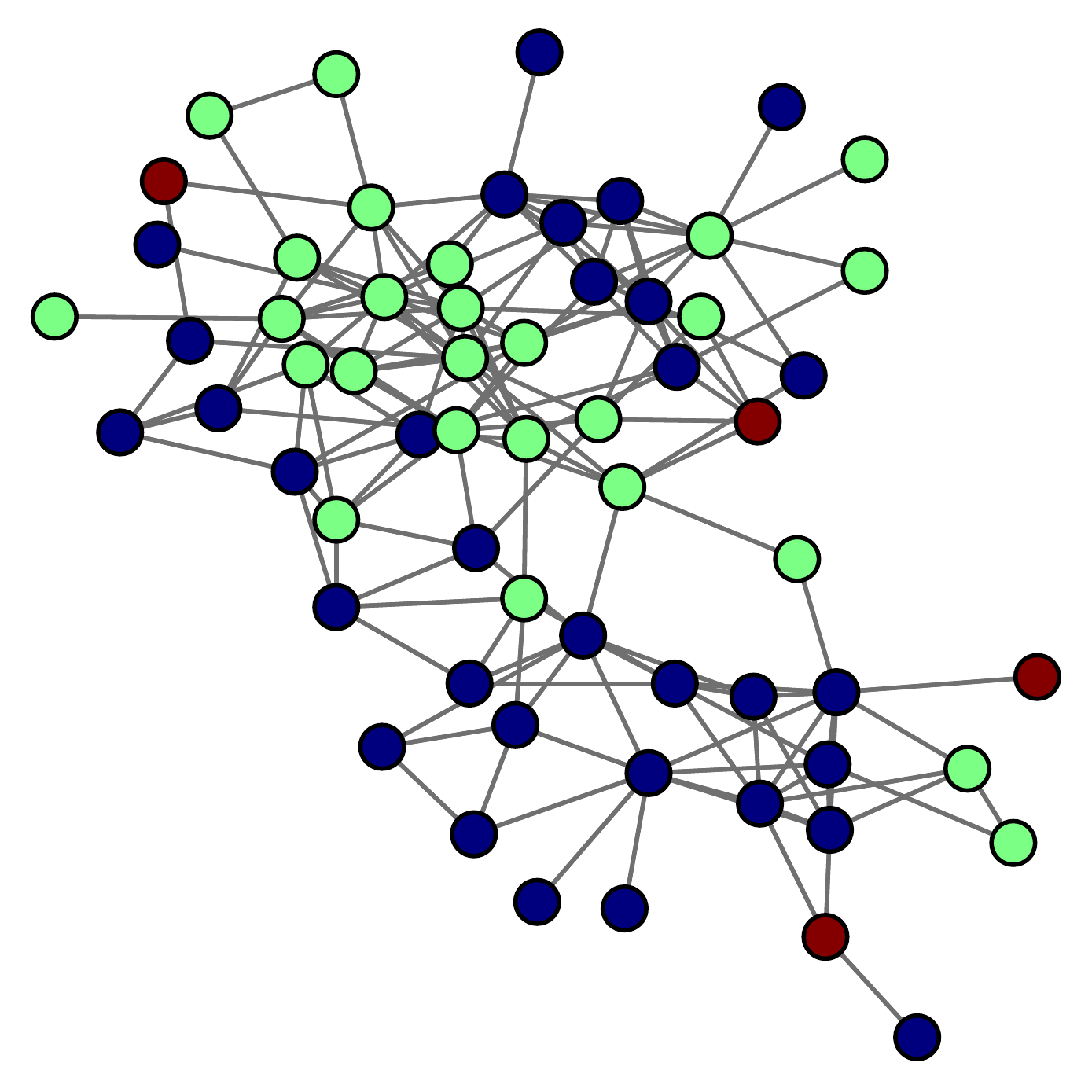}
   \includegraphics[width=0.45\textwidth]{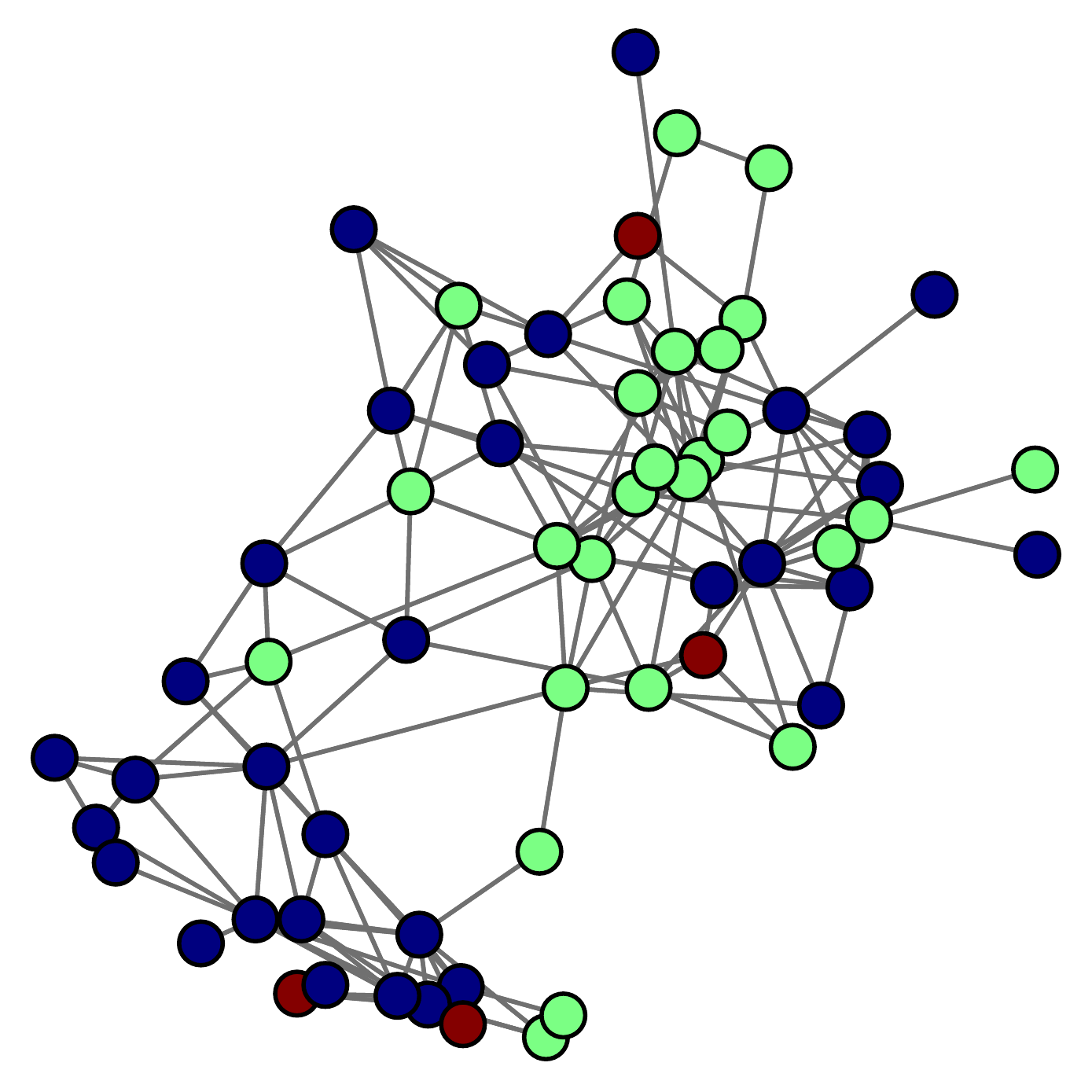}
   \includegraphics[width=0.45\textwidth]{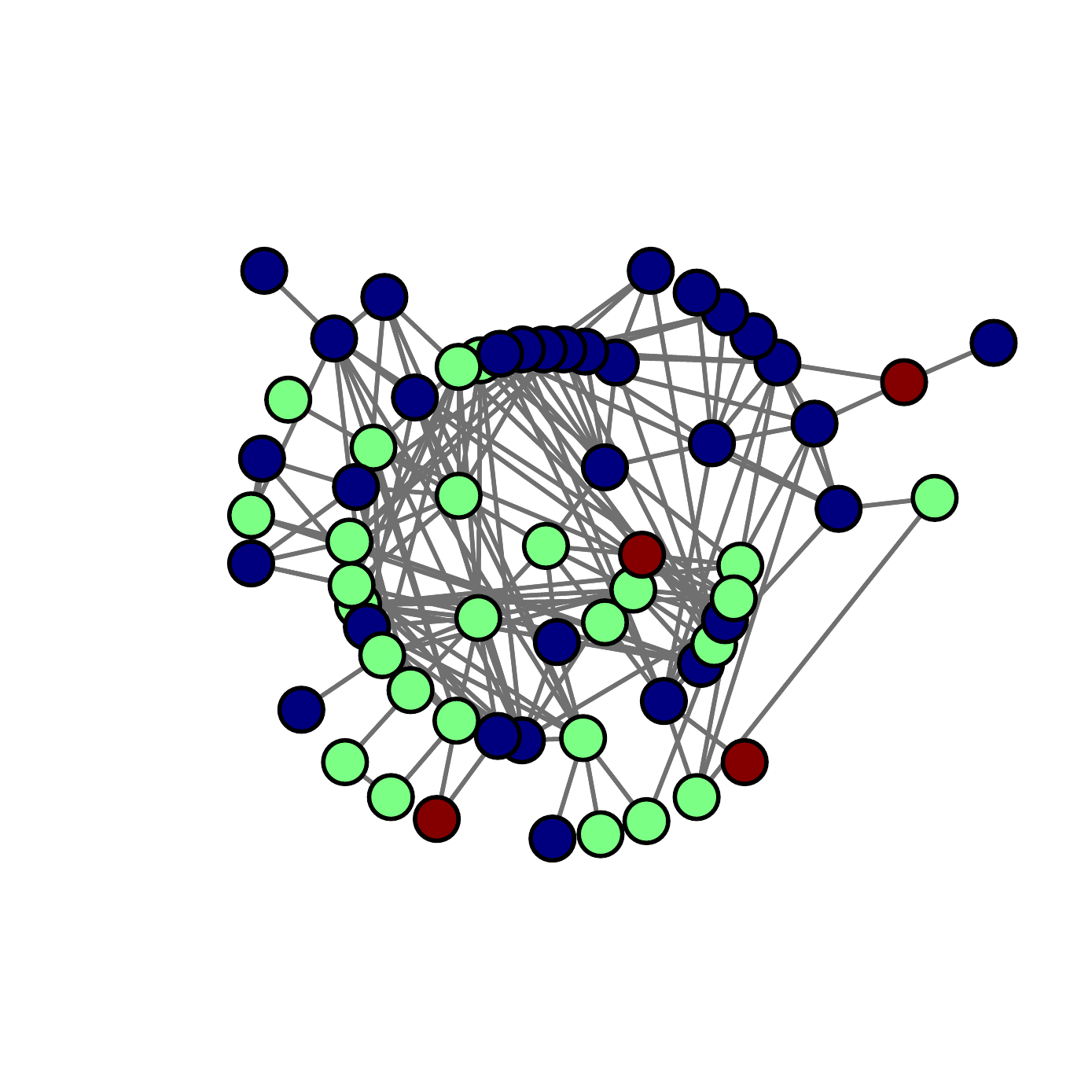}
  \caption{A dolphin social network \cite{lusseau2003bottlenose} represented using different network visualization algorithms: random layout (top left), Kamada--Kawai layout (top right), Fruchterman--Reingold layout (bottom left), and circular layout using average path length (bottom right).}
  \label{fig:visualization}
 \end{figure*}
 
Humans are still better than machines at the recognition of certain patterns when analyzing data in a visual way. Network visualization is a complex task since networks tend to be huge, with thousands nodes and links. NOESIS enables the visualization of networks by providing the functionality needed to render the network and export the resulting visualization using different image file formats.

NOESIS provides different automatic graph layout techniques, such as the well--known Fruchterman--Reingold \citep{fruchterman1991graph} and Kamada--Kawai \citep{kamada1989algorithm} force--based layout algorithms. Force--based layout algorithms assign forces among pairs of nodes and solve the system to reach an equilibrium point, which usually leads to an  aesthetic visualization. 

Hierarchical layouts \citep{tamassia2013handbook}, which arrange nodes in layers trying to minimize edge crossing, are also included. Different radial layout algorithms are included as well \citep{wills1999nicheworks}. These layouts are similar to the hierarchical ones, but arrange nodes in concentric circles. Finally, several regular layouts are included. These layouts are common for visualizing regular networks, such as meshes or stars. 

NOESIS allows tuning the network visualization look and feel. The visual properties of nodes and links can be customized, including color, size, borders, and so on. In addition, visual properties can be bound to static or dynamic properties of the network. For example, node sizes can be bound to a specific centrality score, allowing the visual display of quantitative information.

\section{Network data mining techniques}
Network data mining techniques exist for both unsupervised and supervised settings. NOESIS includes a wide array of community detection methods \citep{lancichinetti2009community} and link prediction techniques \citep{liben2007link}. These algorithms are briefly described below.

\subsection{Community detection}
\begin{figure*}[t]
  \centering
   \includegraphics[width=0.45\textwidth]{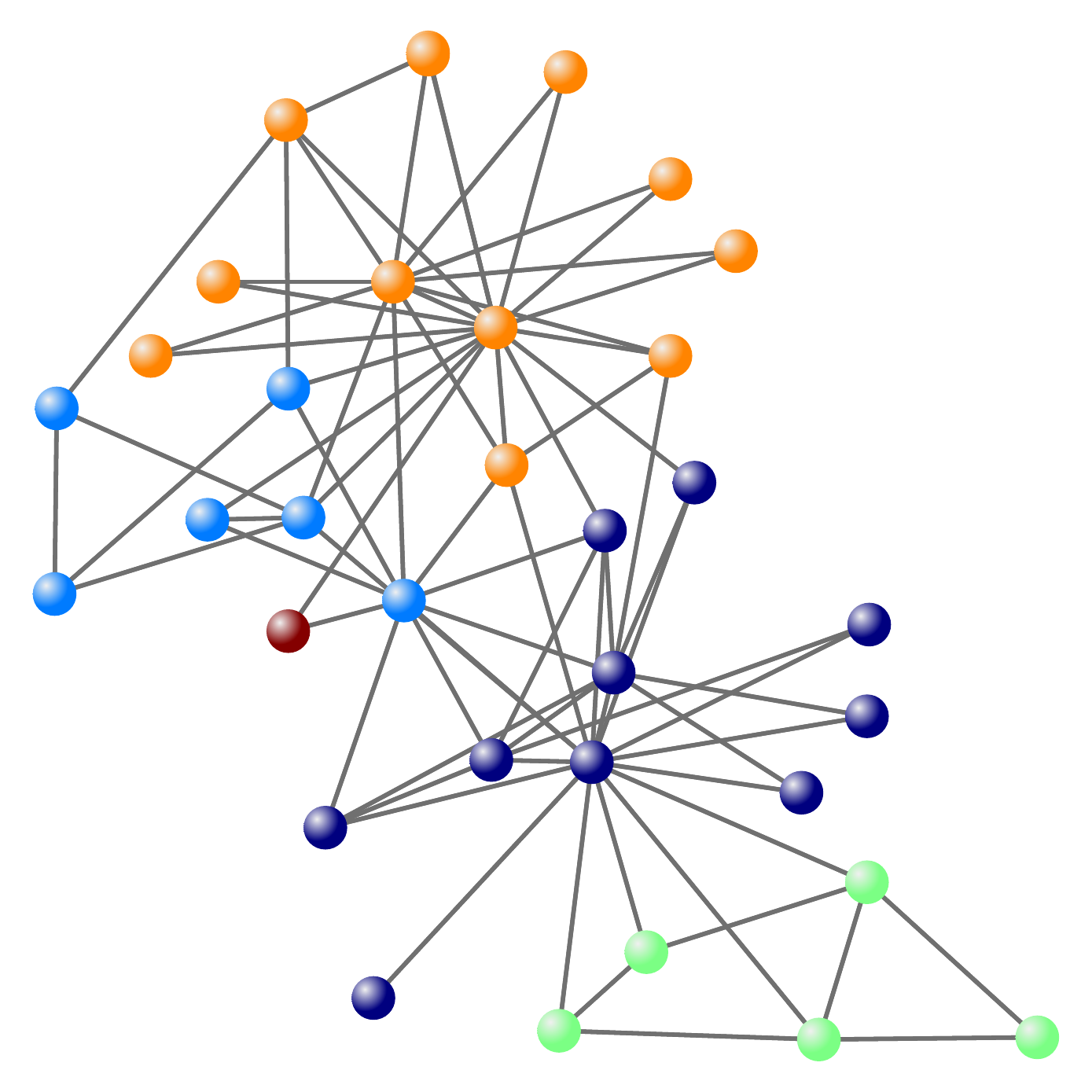}
   \includegraphics[width=0.45\textwidth]{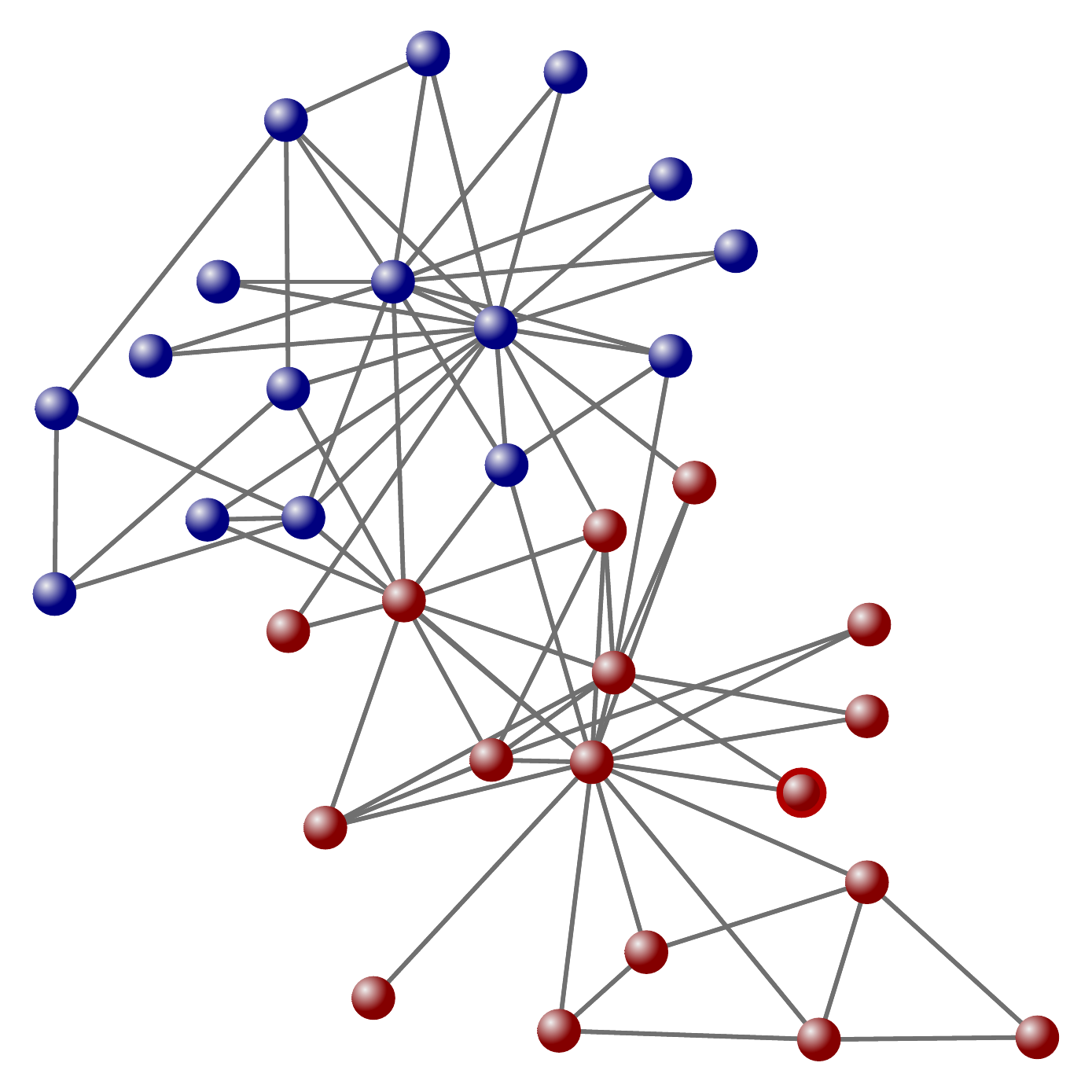}
   \includegraphics[width=0.45\textwidth]{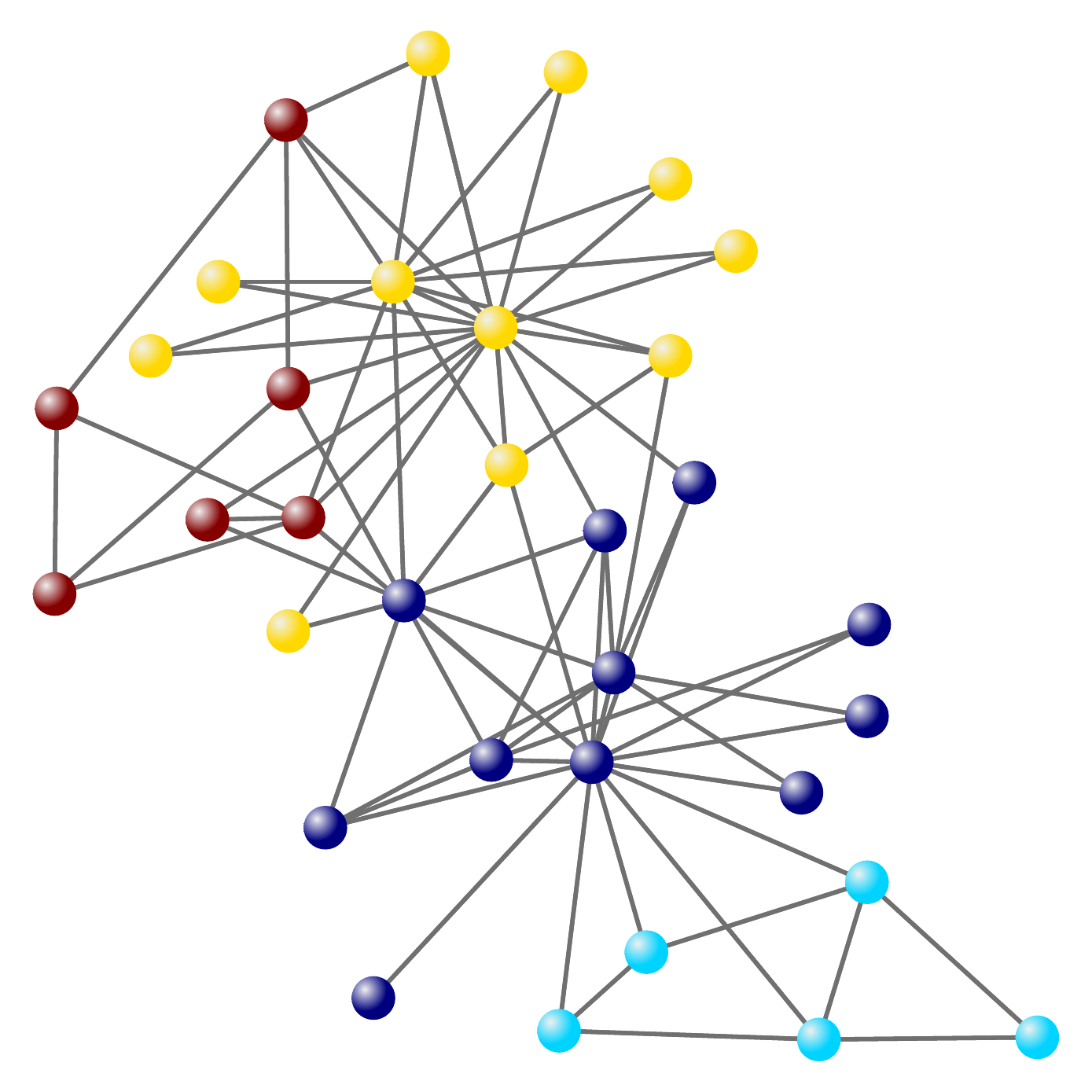}   
   \includegraphics[width=0.45\textwidth]{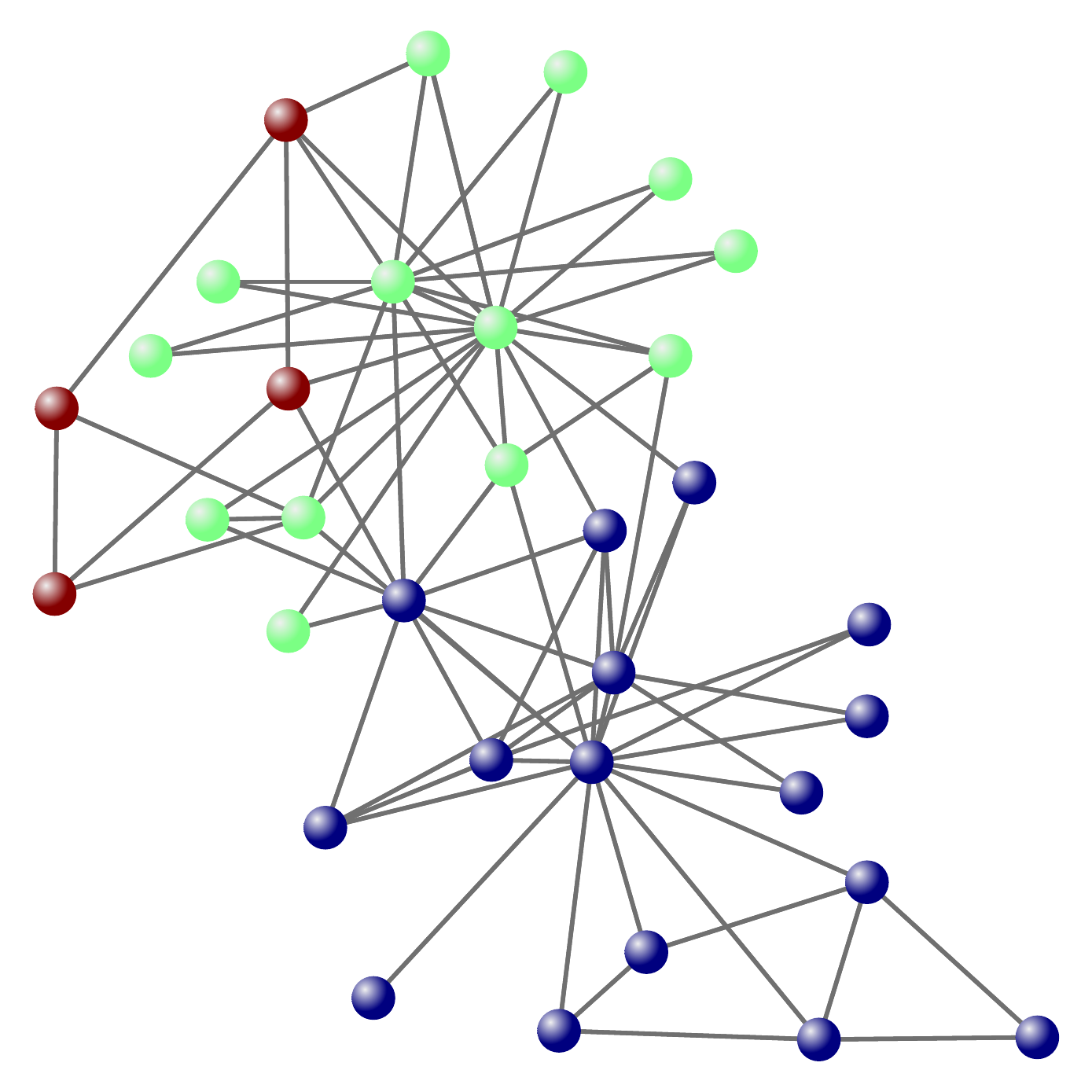}   
  \caption{Community detection methods applied to Zachary's karate club network \cite{zachary1977information}: Fast greedy partitioning (top left), Kernighan-Lin bi-partitioning (top right), average-link hierarchical partitioning (bottom left), and complete-link hierarchical partitioning (bottom right).}
  \label{fig:communitydetection}
 \end{figure*}
 
Community detection can be defined as the task of finding groups of densely connected nodes. A wide range of community detection algorithms have been proposed, exhibiting different pros and cons. NOESIS features different families of community detection techniques and implements more than ten popular community detection algorithms. The included algorithms, their time complexity, and their bibliographic references are shown in Table \ref{tab:clustering_summary}.

\begin{table*}
\centering
\begin{tabular}{|l|l|c|c|}\hline
\textbf{Type} & \textbf{Name} & \textbf{Time complexity} & \textbf{Reference}\\\hline
\multirow{3}{*}{Hierarchical} & Single-link (SLINK) & $O(v^2)$ & \citep{sibson1973slink} \\\cline{2-4}
& Complete-link (CLINK) & $O(v^2 \log v)$ & \citep{defays1977efficient} \\\cline{2-4}
& Average-link (UMPGA) & $O(v^2 \log v)$ & \citep{liu2011webdata} \\\hline
\multirow{2}{*}{Modularity} & Fast greedy & $O(kvd \log v)$ & \citep{newman2004fast} \\\cline{2-4}
 & Multi-step greedy & $O(kvd \log v)$ & \citep{schuetz2008efficient} \\\hline
\multirow{2}{*}{Partitional} & Kernighan-Lin bi-partitioning & $O(v^2 \log v)$ & \citep{kernighan1970efficient} \\\cline{2-4}
 & K-means & $O(kvd)$ & \citep{macqueen1967some} \\\hline
\multirow{3}{*}{Spectral} & Ratio cut algorithm (EIG1) & $O(v^3)$ & \citep{hagen1992new} \\\cline{2-4}
 & Jordan and Weiss NG algorithm (KNSC1) & $O(v^3)$ & \citep{ng2002spectral} \\\cline{2-4}
 & Spectral k-means & $O(v^3)$ & \citep{shi2000normalized} \\\hline
\multirow{1}{*}{Overlapping} & BigClam & $O(v^2)$ & \citep{yang2013overlapping} \\\hline
\end{tabular}
\caption{}{Computational time complexity and bibliographic references for the community detection techniques provided by NOESIS. In the time complexity analysis, $v$ is the number of nodes in the network, $d$ is the maximum node degree, and $k$ is the desired number of clusters.}
\label{tab:clustering_summary}
\end{table*}

NOESIS provides hierarchical clustering algorithms. Agglomerative hierarchical clustering treats each node as a cluster, and then iteratively merges clusters until all nodes are in the same cluster \citep{fortunato2010community}. Different strategies for the selection of clusters to merge have been implemented, including single-link \citep{sibson1973slink}, which selects the two clusters with the smallest minimum pairwise distance; complete-link \citep{defays1977efficient}, which selects the two clusters with the smallest maximum pairwise distance; and average-link \citep{liu2011webdata}, which selects the two clusters with the smallest average pairwise distance.

Modularity-based techniques are also available in our framework. Modularity is a score that measures the strength of particular division into modules of a given network. Modularity--based techniques search for communities by attempting to maximize their modularity score \citep{newman2004finding}. Different greedy strategies, including fast greedy \citep{newman2004fast} and multi-step greedy \citep{schuetz2008efficient}, are available. These greedy algorithms merge pairs of clusters that maximize the resulting modularity, until all possible merges would reduce the network modularity.

Partitional clustering is another common approach. Partitioning clustering decomposes the network and performs an iterative relocation of nodes between clusters. For example, Kernighan-Lin bi-partitioning \citep{kernighan1970efficient} starts with an arbitrary partition in two clusters. Then, iteratively exchanges nodes between both clusters to minimize the number of links between them. This approach can be applied multiple times to subdivide the obtained clusters. K-means community detection \citep{macqueen1967some} is an application of the traditional k-means clustering algorithm to networks and another prominent example of partitioning community detection.

Spectral community detection \citep{fortunato2010community} is another family of community detection techniques included in NOESIS. These techniques use the Laplacian representation of the network, which is a network representation computed by subtracting the adjacency matrix of the network to a diagonal matrix where each diagonal element is equal to the degree of the corresponding node. Then, the eigenvectors of the Laplacian representation of the network are computed. NOESIS includes the ratio cut algorithm (EIG1) \citep{hagen1992new}, the Jordan and Weiss NG algorithm (KNSC1) \citep{ng2002spectral}, and spectral k-means \citep{shi2000normalized}.

Finally, the BigClam overlapping community detector is also available in NOESIS \citep{yang2013overlapping}. In this algorithm, each node has a profile, which consists in a score between $0$ and $1$ for each cluster that is proportional to the likelihood of the node belonging to that cluster. Also, a score between pairs of nodes is defined yielding values proportional to their clustering assignment overlap. The algorithm iteratively optimizes each node profile to maximize the value between connected nodes and minimize the value among unconnected nodes.

In the following example, we show how to load a network from a data file and detect communities with the KNSC1 algorithm using NOESIS:
\begin{verbatim}
FileReader fileReader =
    new FileReader("mynetwork.net");
NetworkReader reader =
    new PajekNetworkReader(fileReader);
Network network = reader.read();
CommunityDetector task =
    new NJWCommunityDetector(network);
Matrix results = task.call();
\end{verbatim}

\subsection{Link scoring and prediction}
Link scoring and link prediction are two closely related tasks. On the one hand, link scoring aims to compute a value or weight for a link according to a specific criteria. Most link scoring techniques obtain this value by considering the overlap or relationship between the neighborhood of the nodes at both ends of the link. On the other hand, link prediction computes a value, weight, or probability proportional to the likelihood of the existence of a certain link according to a given model of link formation.

The NOESIS framework provides a large collection of methods for link scoring and link prediction, from local methods, which only consider the direct neighborhood of nodes, to global methods, which consider the whole network topology. As the amount of information considered is increased, the computational and spatial complexity of the techniques also increases. The link scoring and prediction methods available in NOESIS are shown in Table \ref{tab:linkprediction_summary}.

\begin{table*} 
\centering
\begin{tabular}{|c|l|c|c|}\hline
\textbf{Type} & \textbf{Name} & \textbf{Time complexity} & \textbf{Reference}\\\hline
\multirow{9}{*}{Local} & Common Neighbors count & $O(vd^3)$ & \citep{newman2001clustering}\\\cline{2-4}
& Adamic--Adar score & $O(vd^3)$ & \citep{adamic2003friends} \\\cline{2-4}
& Resource--allocation index & $O(vd^3)$ & \citep{zhou2009predicting} \\\cline{2-4}
& Adaptive degree penalization score & $O(vd^3)$ & \citep{martinez2016adaptive} \\\cline{2-4}
& Jaccard score & $O(vd^3)$ & \citep{jaccard1901etude} \\\cline{2-4}
& Leicht-Holme-Newman score & $O(vd^3)$ & \citep{leicht2006vertex} \\\cline{2-4}
& Salton score & $O(vd^3)$ & \citep{salton1986introduction} \\\cline{2-4}
& Sorensen score & $O(vd^3)$ & \citep{sorensen1948method} \\\cline{2-4}
& Hub promoted index & $O(vd^3)$ & \citep{ravasz2002hierarchical} \\\cline{2-4}
& hub depressed index & $O(vd^3)$ & \citep{ravasz2002hierarchical} \\\cline{2-4}
& Preferential attachment score & $O(vd^2)$ & \citep{barabasi1999emergence} \\\hline
\multirow{7}{*}{Global} & Katz index & $O(v^3)$ & \citep{katz1953new} \\\cline{2-4}
 & Leicht-Holme-Newman score & $O(cv^2d)$ & \citep{leicht2006vertex} \\\cline{2-4}
 & Random walk & $O(cv^2d)$ & \citep{pearson1905problem}\\\cline{2-4}
 & Random walk with restart & $O(cv^2d)$ & \citep{tong2006fast} \\\cline{2-4}
 & Flow propagation & $O(cv^2d)$ & \citep{vanunu2008propagation} \\\cline{2-4}
 & Pseudoinverse Laplacian score & $O(v^3)$ & \citep{fouss2007random} \\\cline{2-4}
 & Average commute time score & $O(v^3)$ & \citep{fouss2007random} \\\cline{2-4}
 & Random forest kernel index & $O(v^3)$ & \citep{chebotarev2006matrix} \\\hline
\end{tabular}
\caption{}{Computational time complexity and bibliographic references for the link scoring and prediction methods provided by NOESIS. In the time complexity analysis, $v$ is the number of nodes in the network, $d$ is the maximum node degree, and $c$ refers to the number of iterations required by iterative global link prediction methods.}
\label{tab:linkprediction_summary}
\end{table*}

Among local methods, the most basic technique is the common neighbors score \citep{newman2001clustering}, which is equal to the number of shared neighbors between a pair of nodes. Most techniques are variations of the common neighbors score. For example, the Adamic--Adar score \citep{adamic2003friends} is the sum of one divided by the logarithm of the degree of each shared node. The resource--allocation index \citep{zhou2009predicting} follows the same expression, but directly considers the degree instead of the logarithm of the degree. The adaptive degree penalization score \citep{martinez2016adaptive} also follows the same approach, yet automatically determines an adequate degree penalization by considering properties of the network topology. Other local measures consider the number of shared neighbors, but normalize their value according to certain criteria. For example, the Jaccard score \citep{jaccard1901etude} normalizes the number of shared neighbors by the total number of neighbors. The local Leicht-Holme-Newman score \citep{leicht2006vertex} normalizes the count of shared neighbors by the product of both neighborhoods sizes. Similarly, the Salton score \citep{salton1986introduction} also normalizes, this time using the square root of the product of both node degrees. The Sorensen score \citep{sorensen1948method} considers the double of the count of shared neighbors normalized by the sum of both neighbors size. The hub promoted and hub depressed scores \citep{ravasz2002hierarchical} normalize the count of shared neighbors by the minimum and the maximum of both nodes degree respectively. Finally, the preferential attachment score \citep{barabasi1999emergence} only considers the product of both node degrees.

 \begin{figure*}[t]
  \centering
   \includegraphics[width=0.45\textwidth]{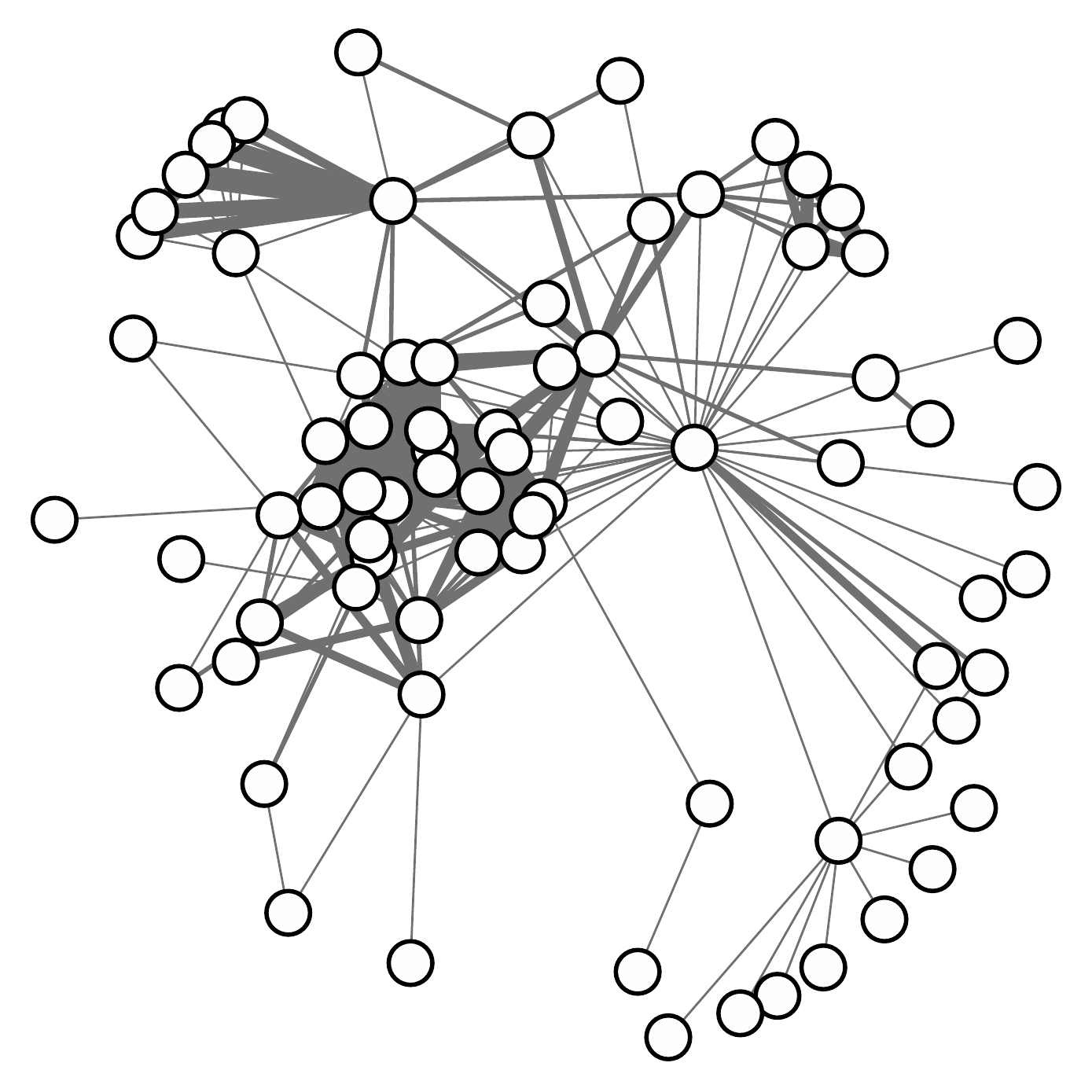}
   \includegraphics[width=0.45\textwidth]{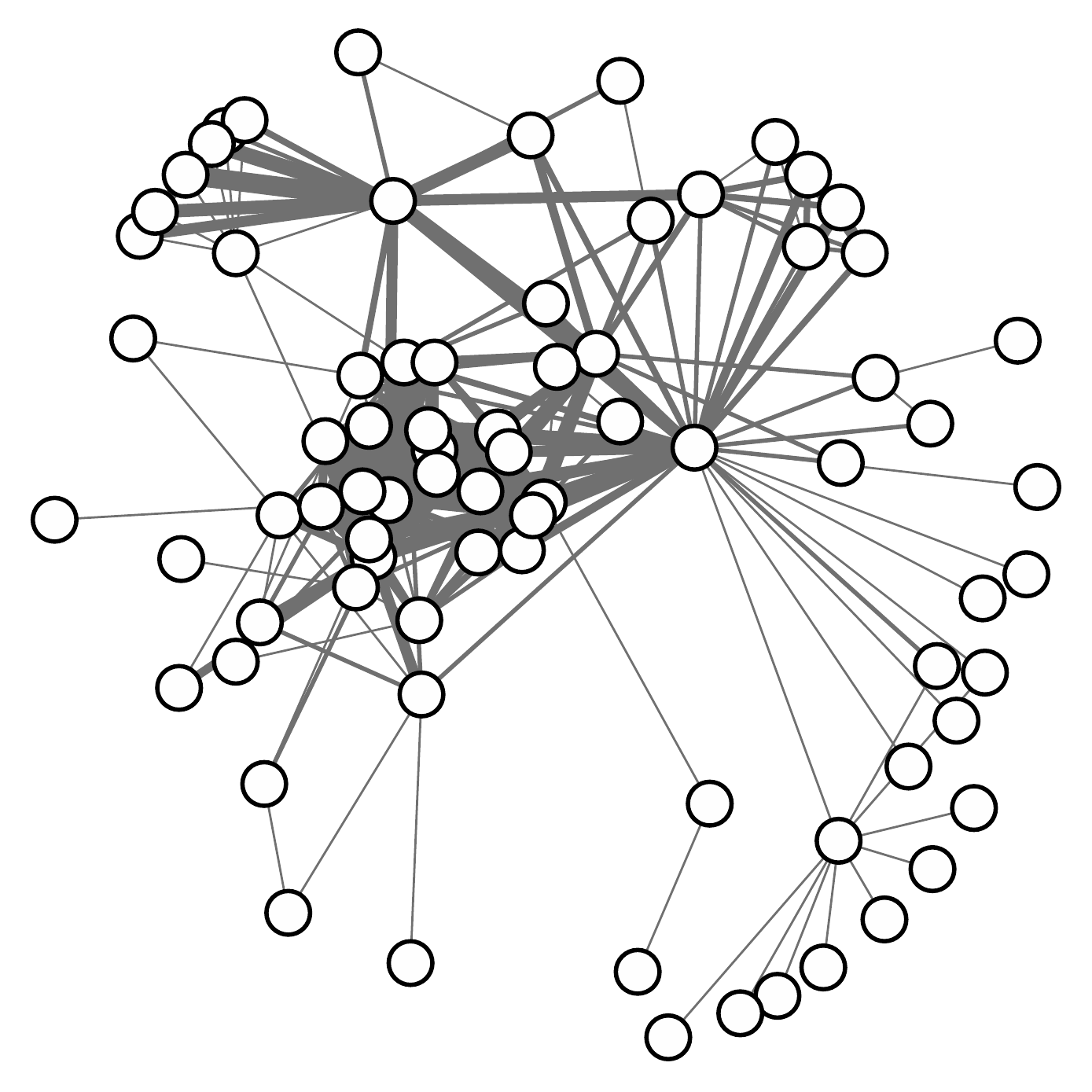}
   \includegraphics[width=0.45\textwidth]{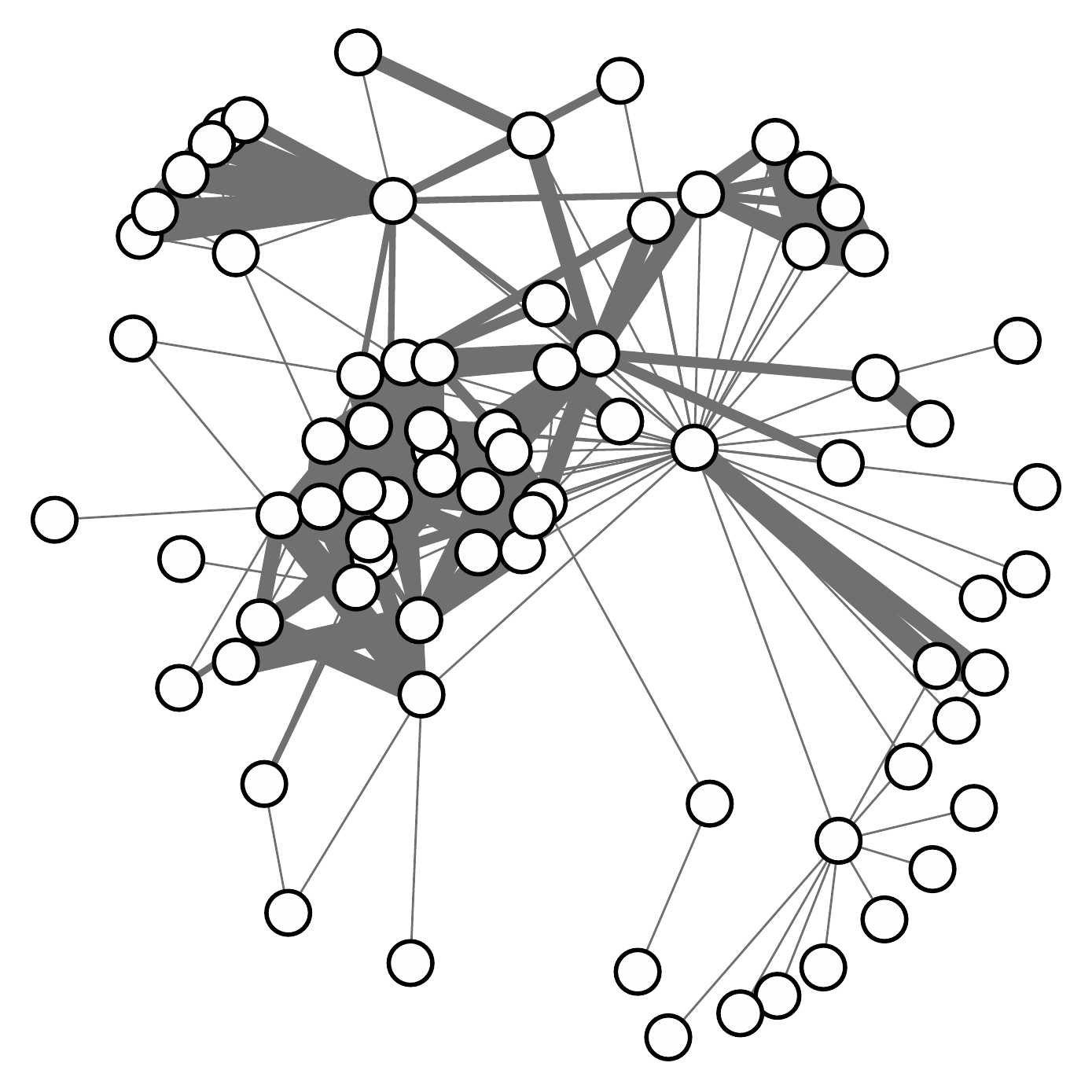}   
   \includegraphics[width=0.45\textwidth]{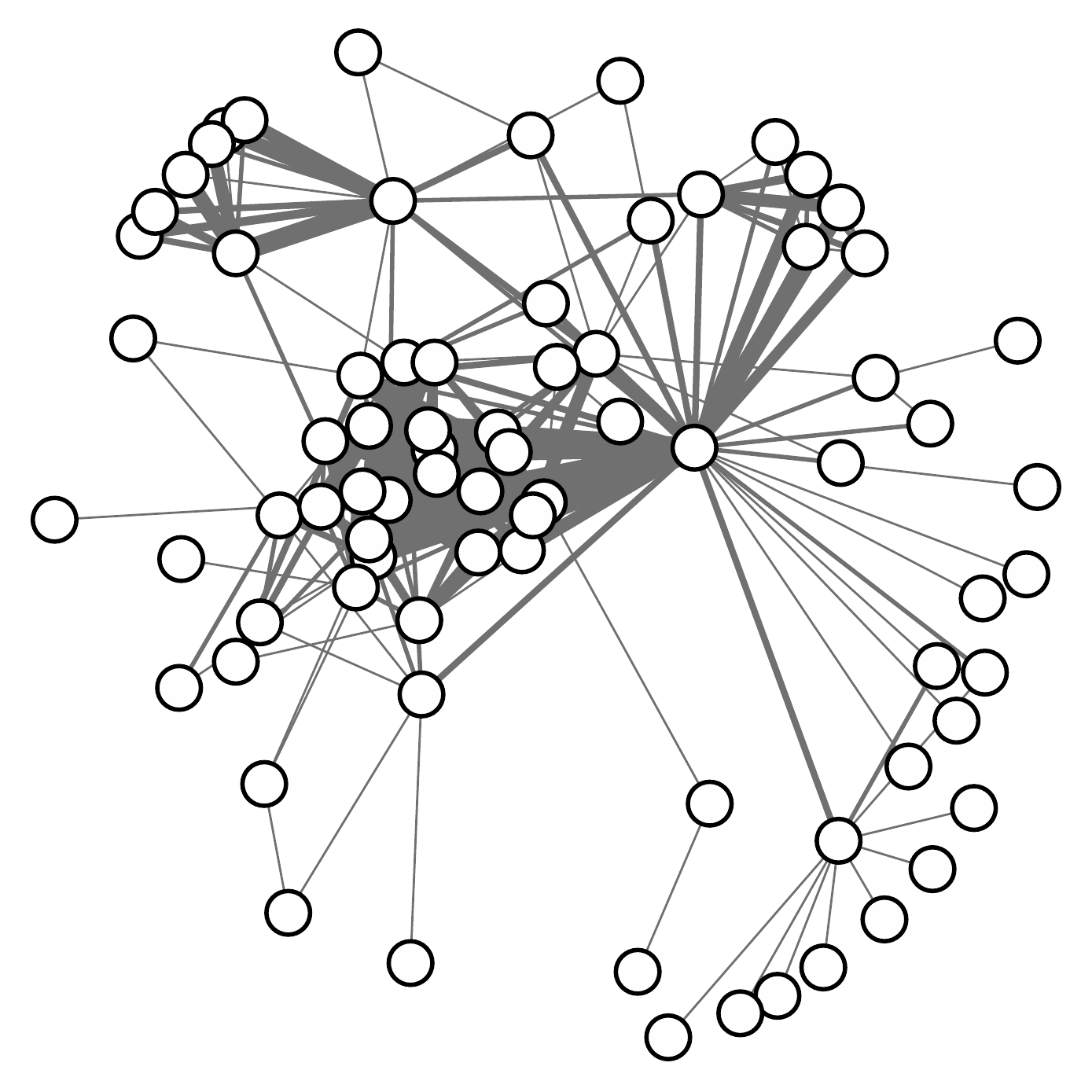}   
  \caption{Different link scoring techniques applied to Les Miserables coappearance network \cite{knuth1993stanford}: common neighbors (top left), preferential attachment score (top right), Sorensen score (bottom left), and Katz index (bottom right). Link width in the figure is proportional to the link score.}
  \label{fig:linkprediction}
 \end{figure*}

Global link scoring and prediction methods are more complex than local methods. For example, the Katz score \citep{katz1953new} sums the influence of all possible paths between two nodes, incrementally penalizing paths by their length according to a given damping factor. The global Leicht-Holme-Newman score \citep{leicht2006vertex} is quite similar to the Katz score, but resorts to the dominant eigenvalue to compute the final result.

Random walk techniques simulate a Markov chain of randomly-selected nodes \citep{pearson1905problem}. The idea is that, starting from a seed node and randomly moving through links, we can obtain a probability vector where each element corresponds to the probability of reaching each node. The classical random walk iteratively multiplies the probability vector by the transition matrix, which is the row-normalized version of the adjacency matrix, until convergence. An interesting variant is the random walk with restart \citep{tong2006fast}, which models the possibility of returning to the seed node with a given probability. Flow propagation is another variant of random walk \citep{vanunu2008propagation}, where the transition matrix is computed by performing both row and column normalization of the adjacency matrix.

Finally, some spectral techniques are also available in NOESIS. Spectral techniques, as we mentioned when discussing community detection methods, are based on the Laplacian matrix. The pseudoinverse Laplacian score \citep{fouss2007random} is the inner product of the rows of the corresponding pair of nodes from the Laplacian matrix. Other spectral technique is the average commute time \citep{fouss2007random}, which is defined as the average
number of steps that a random walker starting from a particular node takes to reach another node for the first time and go back to the initial node. Despite it models a random walk process, it is considered to be a spectral technique because it is usually computed in terms of the Laplacian matrix. Given the Laplacian matrix, it can be computed as the diagonal element of the starting node plus the diagonal element of the ending node, minus two times the element located in the row of the first node and the column of the second one.

Finally, the random forest kernel score \citep{chebotarev2006matrix} is a global technique based on the concept of spanning tree, i.e. a connected undirected sub-network with no cycles that includes all the nodes and some or all the links of the network.  The matrix-tree theorem states that the number of spanning trees in the network is equal to any cofactor, which is a determinant obtained by removing the row and column of the given node, of an entry of its Laplacian representation. As a result of this, the inverse of the sum of the identity matrix and the Laplacian matrix gives us a matrix that can be interpreted as a measure of accessibility between pairs of nodes.

Using network data mining algorithms in NOESIS is simple. In the following code snippet, we show how to generate a Barabási–-Albert preferential attachment network with $100$ nodes and $1000$ links, and then compute the Resource Allocation score for each pair of nodes using NOESIS:

\begin{verbatim}
Network network =
    new BarabasiAlbertNetwork(100, 1000);
LinkPredictionScore method =
    new ResourceAllocationScore(network);
Matrix result = method.call();
\end{verbatim}

\section{Conclusion}
Currently, the NOESIS project comprises more than thirty five thousand lines of code organized in hundreds of classes and dozens of packages. NOESIS relies on a library of reusable components that, with more than forty thousand lines of Java code, provide a customizable collection framework, support for the execution of parallel algorithms, mathematical routines, and the model-driven application generator used to build the NOESIS graphical user interface.

NOESIS can ease the development of applications that involve the analysis of any kind of data susceptible of being represented as a network. NOESIS provides an efficient, scalable, and developer--friendly framework for network data mining, released under a permissive Berkeley Software Distribution free software license. Our framework can be downloaded from its official website at \url{http://noesis.ikor.org}.

\begin{acknowledgments}
The NOESIS project is partially supported by the Spanish Ministry of Economy and the European Regional Development Fund (FEDER), under grant TIN2012--36951, and the Spanish Ministry of Education under the program ``Ayudas para contratos predoctorales para la formaci\'on de doctores 2013'' (grant  BES--2013--064699). We are grateful to Aar\'on Rosas, Francisco--Javier Gij\'on, and Julio--Omar Palacio for their contributions to the implementation of community detection methods in NOESIS.
\end{acknowledgments}

\bibliographystyle{apalike}
\bibliography{arxiv}

\end{document}